\documentstyle[12pt,aaspp4,epsf]{article}
\def\fun#1#2{\lower3.6pt\vbox{\baselineskip0pt\lineskip.9pt
  \ialign{$\mathsurround=0pt#1\hfil##\hfil$\crcr#2\crcr\sim\crcr}}}
\def\lap{\mathrel{\mathpalette\fun <}}
\def\gap{\mathrel{\mathpalette\fun >}}

\def\etal{{\it et al.}}

\received{}
\accepted{ }

\lefthead{Fridman \& Merritt}
\righthead{Orbits in Triaxial Galaxies}

\begin{document}

\title{Periodic Orbits in Triaxial Galaxies with Weak Cusps}

\author{Tema Fridman and David Merritt}
\affil{Department of Physics and Astronomy, Rutgers University,
New Brunswick, NJ 08855}

\begin{abstract}
The orbital structure of triaxial models with weak central 
density cusps, $\rho\propto r^{-\gamma}$, $\gamma<1$, is 
investigated.
The stability of the $x$- (long-) axis orbit -- and hence the
existence of box orbits -- depends sensitively on 
$\gamma$; the range of model shapes for which the $x$-axis orbit 
is stable becomes progressively smaller as $\gamma$ approaches 
one.
The banana and fish boxlets in the $x-z$ (long axis--short axis) 
plane are stable over a wide range of model parameters.
The boxlets in the $x-y$ and $y-z$ planes are generally vertically 
unstable.

\end{abstract}

\section {Introduction}

The phase space of St\"ackel potentials is completely occupied by 
four major families of regular orbits: the boxes, and the three 
families of tubes (\cite{kuz73}; \cite{del85}).
Roughly the same is true in many nonrotating triaxial potentials 
corresponding to mass models with smooth cores: most of the orbits 
have shapes that can be identified with one of the four families 
of regular orbits in St\"ackel potentials (\cite{sch79}).
However the situation is very different in triaxial models with 
divergent central densities, or cusps.
While the tube orbits are not strongly affected by cusps, 
the boxlike orbits -- orbits with stationary points and filled 
centers -- are often rendered stochastic, except when they 
are associated with a resonant orbit that avoids the 
center (\cite{geb85}; \cite{les92}; \cite{mef96}; \cite{pal97}).

Kormendy pointed out already in 1985 that many early-type 
galaxies and bulges have central brightness profiles that deviate 
slightly but systematically from that of an isothermal core.
The significance of this deviation was not recognized for 
ten more years, however, due to an optical illusion associated with 
projection onto the plane of the sky.
A luminosity density profile that varies as $r^{-\gamma}$ at small 
radii generates a power-law cusp in projection only if $\gamma>1$.
When $\gamma=1$, the surface brightness exhibits a
curving, logarithmically-divergent central profile
(e.g. \cite{deh93}, Fig. 1), and for $\gamma<1$ the central
surface brightness is finite.
The observed brightness profile of a galaxy like M87, which has
a cusp with $\gamma\approx 0.8$ (\cite{lau92}), differs only subtly from 
that of a galaxy with an isothermal core (\cite{kor85}).
This is true in spite of the fact that the cusp in M87 is
well resolved from the ground.

Thus even if galaxies were distributed uniformly over
$\gamma$, their central brightness profiles would 
appear to fall into one of two distinct categories: the ``cores''
($\gamma\leq 1$) and the ``power-laws'' ($\gamma>1$).
Just such a dichotomy was proposed following the first 
nuclear brightness measurements from HST (\cite{fer94}; 
\cite{lau95}).
Merritt \& Fridman (1995) suggested that the 
centers of {\it all} early-type galaxies, including the core
galaxies, might contain power law cusps in the space density.
They used a nonparametric deprojection algorithm to 
confirm this hypothesis for six galaxies observed with HST.
Gebhardt et al. (1996) subsequently reanalyzed the full sample 
of 42 early-type galaxies observed by Lauer et al. (1995) 
and verified the power-law nature of the deprojected cusps in each case.
They also confirmed a tendency, noted earlier by Kormendy (1985), 
for the brightest galaxies to have the shallowest cusps.

Most of the theoretical work on orbital motion in triaxial potentials
has focused on mass models with cores, $\gamma=0$, or with steep cusps, 
$\gamma=2$.
But galaxies with cores probably do not exist; and only the faintest 
ellipticals, $M_v\gap -19$, have cusps that are predictably as 
steep as $r^{-2}$.
Here we present the first detailed study of periodic orbits in 
triaxial models with weak cusps, $0\le\gamma<1$, characteristic of 
bright elliptical galaxies.
The value $\gamma=1$ is a natural one for separating ``weak'' 
from ``strong'' cusps, since a central density that increases 
more rapidly than $r^{-1}$ implies a divergent central force.
For instance, a spherical galaxy with Dehnen's (1993) density
law
\begin{equation}
\rho(r) = {(3-\gamma)M\over 4\pi a^3}\left({r\over 
a}\right)^{-\gamma}
\left(1+{r\over a}\right)^{-(4-\gamma)}
\end{equation}
has a gravitational force
\begin{equation}
-{\partial\Phi\over\partial r} = -{GM\over a^2}\left({r\over
a}\right)^{1-\gamma}\left(1+{r\over a}\right)^{\gamma-3}.
\end{equation}
Dehnen's law, with $\gamma=2$ (\cite{jaf83}) and $\gamma=1$ (\cite{her90}), 
has been shown to accurately describe the brightness profiles of a 
number of early-type galaxies.
The triaxial generalization of Dehnen's law,
\begin{equation}
\rho(m)={(3-\gamma)\over 4\pi abc} m^{-\gamma}(1+m)^{-(4-\gamma)}, 
\end{equation}
\begin{equation}
m^2={x^2\over a^2}+{y^2\over b^2}+{z^2\over c^2},\ \ \ a\ge b\ge c\ge 
0,
\end{equation}
was presented by Merritt \& Fridman (1996), who derived 
expressions for the gravitational potential, force and force 
gradients for $0\le\gamma\le 2$.

Our primary concern here is with the closed ``boxlets'', the 
low-order resonant orbits that act as parents of the boxlike 
orbits in non-integrable triaxial potentials.
Most important is the $x$- 
(long-) axis orbit, which generates the box orbits in 
St\"ackel potentials.
Its instability usually occurs through the bifucation of a 
$2:1$ resonant orbit, or ``banana.'' 
(We adopt Miralda-Escud\'e \& Schwarzschild's (1989) scheme for 
labeling the closed boxlets; their Figure 4 illustrates the 
different types.)
Higher-order resonances produce the  ``fish'' ($3:2$) and ``pretzel'' 
($4:3$) orbits.
Boxlets corresponding to each resonance are expected to exist in 
all three of the principal planes of a triaxial model, at least at 
certain energies and for certain values of the model axis ratios.
However little is known about the dependence of boxlet stability 
on model parameters.
In particular, the vertical stability of the closed boxlets has 
been very little explored.

In the present paper, we locate and test the stability of the 
axial orbits, the bananas, and the fish as a function of the three 
parameters that define Dehnen's models: the cusp slope $\gamma$; the 
short-to-long axis ratio $c/a$; and the degree of triaxiality, expressed 
in terms of $T = (a^2-b^2)/(a^2-c^2)$.
$T=0$ corresponds to oblate-axisymmetry, and $T=1$ to 
prolate-axisymmetry.
For each set of model parameters, we present results as a function of 
energy; thus we explore a four-dimensional parameter space for 
each of the orbit families.
A total of approximately $3.5\times 10^4$ orbits were 
investigated.
The numerical scheme for evaluating the stability of the 
boxlets is presented in \S 2.
The results for the axial orbits, the bananas and the fish are 
presented in \S 3, 4 and 5 respectively.
Our results are summarized in \S 6.

There is increasingly strong evidence for dark mass concentrations,
possibly supermassive black holes, at the centers 
of many early-type galaxies (\cite{kor95}).
We ignore the (possibly substantial) effect that such singularities would 
have on the behavior of boxlike orbits: partly because the 
universality of central black holes has not yet been established; 
and partly because the behavior of orbits in triaxial galaxies 
with cusps provides an essential first step toward understanding 
the additional effects of a central black hole.

\section{Linear Stability Analysis}
We applied a numerical version of the standard Floquet-Liapunov theory 
(\cite{flo83}; \cite{lia92}) to evaluate the stability of the 
lowest-order periodic orbits in Dehnen's model potential.
Our approach was similar to that of earlier studies (e.g. 
\cite{mag82}; \cite{pfe84}), with one difference.
Instead of integrating a set of orbits whose initial conditions 
were offset by small, finite amounts from those of the closed orbit,
we integrated the linearized equations corresponding to a set of
infinitesimal perturbations about the closed orbit.
Our results are therefore strictly independent of the 
numerical amplitude of the perturbation.

Let ${\bf X}(t)$, ${\bf V}(t)$ be the parametric representation of a
closed orbit.
The equations of motion are
\begin{equation}
\frac{d{\bf X}(t)}{d t}={\bf V}(t), \ \ \  
\frac{d{\bf V}(t)}{d t}=- \nabla \Phi ({\bf X}).
\label{motion}
\end{equation}
A nearby orbit has coordinates ${\bf X}(t)+{\bf x}(t)$, 
${\bf V}(t) + {\bf v}(t)$.
For small $\bf x$ and $\bf v$, the equations of motion are
\begin{eqnarray}
\frac{d({\bf X} + {\bf x})}{dt} & = & {\bf V} + {\bf v} , \\
\frac{d({\bf V} + {\bf v})}{dt} & = & - \nabla \Phi ({\bf X} + {\bf 
x}) \approx - \nabla \Phi (\bf X)+({\bf x} \nabla)(\nabla \Phi ({\bf 
X})) \nonumber.
\label{linmotion}
\end{eqnarray}
Subtracting equations (\ref{motion}) from equations (\ref{linmotion}),
we have
\begin{equation}
\frac{d{\bf x}}{dt}={\bf v} , \ \ \ 
\frac{d{\bf v}}{dt}=-({\bf x} \nabla)(\nabla \Phi ({\bf X})) ,
\end{equation} 
or
\begin{eqnarray}
 \frac{d}{dt} \left( \begin{array}{l}  x\\y\\z\\v_x\\v_y\\v_z 
\end{array} \right) = - \left( \begin{array}{cccccc}
0&0&0&-1&0&0\\0&0&0&0&-1&0\\0&0&0&0&0&-1\\
\Phi _{xx}&\Phi_{xy}&\Phi_{xz}&0&0&0\\
\Phi _{xy}&\Phi_{yy}&\Phi_{yz}&0&0&0\\
\Phi _{xz}&\Phi_{yz}&\Phi_{zz}&0&0&0 \end{array} \right)
\left( \begin{array}{l} x\\y\\z\\v_x\\v_y\\v_z  
\end{array} \right) .
\label{matrix}
\end{eqnarray}
The second derivatives of the potential, $\Phi_{ij}$, are taken 
along the closed orbit whose period is $P$.
Expressions for the $\Phi_{ij}$ are given by Merritt \& Fridman 
(1996).
The stability of the closed orbit is then determined by the 
eigenvalues of the monodromy matrix $\bf W$, where
\begin{equation}
{\bf x}(t+P)={\bf W}{\bf x}(t)
\end{equation}
and $\bf x$ has been redefined to include velocities.
We computed the elements of $\bf W$ by numerically integrating the 
six linearized equations of motion (\ref{matrix}), starting from 
unit perturbations in each of the coordinates.
A 7/8th order, variable timestep integrator -- the routine DOPRI8 
of Hairer et al. (1987) -- was used.
Initial conditions for the non-axial closed orbits were found by 
an iterative Newton method, and precise orbital periods were calculated 
using the scheme of H\'enon (1982).

Two of the eigenvalues of $\bf W$ are guaranteed to be unity; 
these correspond to eigenvectors lying along the closed orbit.
The remaining four eigenvalues satisfy $\lambda_1=1/\lambda_2$, 
$\lambda_3=1/\lambda_4$; the stability of the closed orbit is 
determined by the parameters $b_1=-(\lambda_1+\lambda_2)$, 
$b_2=-(\lambda_3+\lambda_4)$, with $|b_i|\ge 2$ denoting instability 
(\cite{bro69}).
For orbits that lie in a symmetry plane, $b_1$ and $b_2$ 
correspond to perturbations in the orbital plane and 
perpendicular to the orbital plane, or vice versa.
For the simplest orbits, the full set of six perturbation equations 
is not required; for instance, the stability of the axial orbits 
can be evaluated using just four independent perturbations.

Axial orbits presented one complication, however.
While the gravitational force $-\nabla\Phi$ is everywhere finite 
in Dehnen models with $\gamma<1$, the second derivatives of the 
potential diverge at the origin as $\Phi_{ii}\propto 
|x_i|^{-\gamma}$.
To avoid this divergence, analytical expressions were derived 
for the evolution of the perturbations very near the origin, 
allowing this region to be excluded from the numerical integrations.

The accuracy of the algorithms was checked in a number of ways.
The monodromy matrix and its eigenvalues were checked to see that 
they accurately satisfied all of their known, exact properties, 
e.g. $\det {\bf W}=1$, $\lambda_1=1/\lambda_2$, etc.
Integrations using an algorithm based on finite-amplitude perturbations
were carried out for some orbits and compared to the results of 
the linearized code.
Finally, numerical surfaces of section were constructed for a few 
models and compared to the results from the stability 
calculations.

All orbits in a given model were assigned energies from a discrete
grid, consisting of 20 energy values that corresponded to the values of the 
potential on the $x$-axis of a set of ellipsoidal shells -- with 
the same axis ratios as the density -- that divide the model into 
21 sections of equal mass.
Thus, shell 1 encloses $1/21$ of the total mass, shell 2 encloses 
$2/21$, etc.
Henceforth units are adopted such that $G=M=a=1$.

\section {Axial Orbits}

The long- ($x$-) axis orbit is stable at all energies in integrable 
triaxial potentials and acts as the parent of the box orbits 
(\cite{kuz73}; \cite{del85}).
In more general triaxial potentials, like the ones considered 
here, the $x$-axis orbit becomes unstable at high energies
when its frequency of oscillation falls to $1/2$ the average 
oscillation frequency of a perturbation in the direction of the 
short or intermediate axis (\cite{mis89}).
A bifurcation then occurs, with the $2:1$ $x-z$ or $x-y$ banana 
orbit branching off.
A similar bifurcation occurs from the intermediate ($y$-) axis 
orbit, producing the $y-z$ banana.

Nearer the center, the $y$- and $z$-axis orbits become unstable at 
the bifurcation points of the $1:1$ loop orbits.
In models with a harmonic core, the loop orbits first appear
just outside of the core (\cite{mez82}).
The $x-y$ loop is the parent of the short-axis tubes,
and the $y-z$ loop is the parent of the long-axis tubes.
The $x-z$ loop is generally unstable and generates no families 
of regular orbits (\cite{hes79}).
In models without cores, the loops and their associated tube 
orbits may exist all the way into the center.

Figures 1 and 2 display the stability diagrams for the $x$- 
and $y$- axis orbits in Dehnen's potential, for three values of 
$T$ and four values of $\gamma$; a total of 4104 orbits are 
displayed in each figure.
Four cases are distinguished: stability ($\bullet$); instability 
in the direction of the longer ($+$) or shorter ($\times$) of the 
two remaining axes; and instability in both directions ($\cdot$). 

The $x$-axis orbit is stable at low energies in all models that 
are not too highly elongated.
As $\gamma$ increases above zero, the range of model shapes for 
which the $x$-axis orbit is stable to the $x-z$ banana 
bifurcation becomes narrower, eventually 
including only nearly spherical models (Figure 3). 
For instance, when $\gamma=0.9$, the $x$-axis orbit is stable
at the half-mass radius for $c/a\gap 0.8$.
Thus, bona-fide box orbits (which require for their existence a 
stable long-axis orbit) will occur at a wide range of 
energies only in models that are nearly spherical, that have 
very weak cusps, or both. 
Figure 4 illustrates how the bifurcation of the $x-z$ banana 
orbit takes place at successively lower energies as $\gamma$ is 
increased.

As $c/a$ is decreased still more, a second $2:1$ bifurcation from the 
$x$-axis orbit sometimes occurs in the direction of the 
intermediate axis, giving rise to the $x-y$ bananas.
Following this bifurcation, the $x$-axis orbit is unstable in 
both directions.
As Figure 1 shows, the $x-y$ banana bifurcation occurs most readily in 
nearly prolate models (e.g. $T=0.8$) in which the $y$ and $z$ axes 
are nearly equal in length.
For nearly oblate models, e.g. $T=0.2$, the $x$ and $y$ axes are too 
similar in length for the $x-y$ frequency ratio to ever reach the required 
value of $2:1$, and the $x-y$ banana does not appear.

Interestingly, the $x$-axis orbit can again become stable
when $c/a$ is made smaller still -- typically less than about 
0.45.
The return to stability coincides with the appearance of the 
$x-z$ ``anti-banana'' orbit, a $2:1$ resonant orbit that passes 
through the center (\cite{mis89}).
In nearly oblate models, this return to stability in the $x-z$ 
plane causes the $x$-axis orbit to become stable in both 
directions, at least over a limited range in energy and in $c/a$ (Figure 1).
The axial orbit becomes unstable once again at still lower values of $c/a$, 
through the appearance of a $3:1$ resonance in the direction of 
the $z$ axis (Figure 5).
In nearly prolate models, where the $x$-axis orbit is also unstable 
to perturbations in the $x-y$ plane, the appearance of the $x-z$ 
anti-banana orbit leaves the $x$-axis orbit still unstable to 
perturbations in the $y$ direction (Figure 1).

At very low values of $c/a$, the stability of the $x$-axis orbit 
varies in a rapid and complex manner with energy and with $\gamma$, 
as resonances of higher and higher order occur.

The $y$-axis orbit (Figure 2) is unstable at almost all energies in 
the full range of models investigated here.
Instability in the $x$-direction is expected due to the
$1:1$ resonance that generates the $x-y$ loops; these 
orbits -- and their associated family, the short-axis tubes -- are 
present almost universally in Dehnen's models.
As $c/a$ decreases, the $y$-axis 
orbit also becomes unstable through the $2:1$ bifurcation that 
produces the $y-z$ bananas; the $y$-axis orbit is then unstable 
in both directions.
At still greater model elongations, the $y$-axis 
orbit returns briefly to stability in the $z$-direction through the 
appearance of the $y-z$ anti-banana.

The stability diagram of the $z$-axis orbit is not given here.
This orbit was found to be unstable to perturbations in both the $x$- 
and $y$-directions for almost all values of the model parameters, 
due to the $1:1$ bifurcations that produce 
the $x-z$ and $y-z$ loops (\cite{gos81}).
Thus, the long-axis tube orbits -- which are parented by the 
$y-z$ loop orbits -- are almost always present in these models.
The only exceptions occur at low energies in models with $c/a\lap 
0.2$ and $\gamma\lap 0.5$, where the ratio of oscillation 
frequencies in the $z$- and $x$-directions is too great for the 
$1:1$ resonance to occur.
The $z$-axis orbit is not the parent of any banana family.

In summary: only the $x$-axis orbit is stable over an appreciable 
range of model parameters.
Instability first appears through the $x-z$ banana bifurcation, 
which occurs at progressively larger values of $c/a$ as $\gamma$ 
is increased.
The $x$-axis orbit returns to stability for a narrow range of 
$c/a$ values, typically around $0.2-0.4$, through the bifurcation 
of the $x-z$ anti-banana.

\section {Banana Orbits}

Banana orbits first appear as $2:1$ bifurcations from the $x$-axis 
orbit ($x-z$ and $x-y$ bananas) or the $y$-axis orbit ($y-z$ banana).
Stability diagrams for the $x-z$ and $x-y$ bananas are 
shown in Figures 6 and 7.
Once these orbits bifurcate from the axial orbits, they continue 
to exist (though not always stably) at all higher energies and 
all lower values of $c/a$, for any given value of $\gamma$ and $T$.

The $x-z$ banana exists and is stable over the widest range of 
model parameters (Figure 6).
It becomes unstable only in nearly-prolate models, through a vertical 
$2:1$ bifurcation; for example, when $T=0.8$, instability first occurs 
when $c/a$ drops below $\sim 0.35$.
Thus, in virtually all models with axis ratios comparable to 
those of real elliptical galaxies, either the $x$-axis orbit or
the $x-z$ banana orbit is stable and can act as the parent of a family of 
regular boxlets.

The $x-y$ banana exists only in relatively prolate and
elongated models, in which the $x-y$ frequency ratio can exceed 
$2:1$ (Figure 7).
For moderate values of $c/a$, however, the $x-y$ banana is 
generally vertically unstable; its instability follows closely that 
of the $x-$axis orbit in the $z$-direction (Figure 1).
For smaller values of $c/a$, e.g. $c/a\lap 0.5$ for $T=0.8$, 
the $x-y$ banana returns to stability via a $2:1$ bifurcation
at approximately the same 
model parameters where the $x-$axis orbit also becomes stable in 
the $z-$direction (Figure 8).

The $y-z$ banana was found to be almost universally unstable in the 
vertical ($x$) direction, mimicking the $x$-instability of the $y$-axis 
orbit.

The regular orbits associated with the $x-z$ bananas are known to 
be heavily populated in self-consistent triaxial models with 
cusps (e.g. \cite{sch93}; \cite{mef96}).
However their usefulness is limited by the fact that the banana orbits 
become more strongly curved as $c/a$ decreases, so that in 
sufficiently elongated models, the boxlets associated with the 
$x-z$ bananas are all rounder than the model figure.
Figure 9 shows the value of $c/a$ at which the $x-z$ banana has 
an elongation -- defined as $|z_{max}/x_{max}|$, the ratio of 
its extent in the $z$ and $x$ directions -- that just matches 
that of the model.
Figure 9 suggests that the boxlets associated with the $x-z$ 
banana will be useful in reconstructing triaxial models for
$c/a\gap 0.4$.

In summary: the $x-z$ banana exists and is stable for the widest 
range of model parameters.
It is slender enough to match the shape of the model when $c/a\gap 
0.4$.
The $x-y$ and $y-z$ bananas either do not exist, or are vertically 
unstable, in all models with moderate values of $c/a$; the $x-y$ 
banana returns to stability only in highly elongated, nearly prolate models. 

\section {Fish Orbits}

Fish orbits first appear as $3:2$ bifurcations from the $x$-axis 
orbit ($x-z$ and $x-y$ fish) or the $y$-axis orbit ($y-z$ fish).
Stability diagrams for the $x-z$ and $x-y$ fish are 
shown in Figures 10 and 11.
Once these orbits bifurcate from the axial orbits, they continue 
to exist at all higher energies and all lower values of $c/a$, 
for any given value of $\gamma$ and $T$.

The $x-z$ fish exists and is stable in models with moderate values of 
$c/a$ (Figure 10).
In nearly oblate models, the $x-z$ fish first becomes unstable to 
perturbations in the orbital plane, while for strongly triaxial 
and prolate models, instability first appears in the vertical 
($y$) direction (Figure 12).
However this orbit exists stably over a substantially narrower range of 
model parameters than the $x-z$ banana.

The $x-y$ fish is only important in strongly prolate models 
(Figure 11).
In strongly triaxial and oblate models, it either does not exist, 
or is generally unstable to vertical ($z$) perturbations.

The $y-z$ fish was found to be unstable to vertical perturbations 
for almost all values of the model parameters.

In summary: the $x-z$ fish is important in moderately flattened, 
oblate/triaxial models, while the the $x-y$ fish is important in 
highly prolate models.
In strongly triaxial models, instability in both families first 
occurs in the vertical direction.
Neither orbit exists stably for as wide a range of model 
parameters as the $x-z$ banana.

\section {Discussion}

Our results highlight the sensitivity of boxlike orbits to 
small changes in the central density structure of a triaxial galaxy.
The range of model shapes for which the $x$-axis orbit -- the 
progenitor of the box orbits -- is stable becomes rapidly
narrower as the cusp slope $\gamma$ approaches one.
For $\gamma\gap 0.8$, only triaxial models with moderate axis 
ratios, $c/a\gap 0.75$, contain stable long-axis orbits at most 
energies.
Regular box orbits may nevertheless be present in the majority 
of bright elliptical galaxies, since the distribution of intrinsic 
shapes of these galaxies is strongly peaked at $c/a\approx 0.8$ 
(\cite{trm96}), and many ellipticals have cusp slopes 
$\gamma\lap 0.8$ (\cite{mef95}; \cite{geb96}).

Of the remaining families of boxlets investigated here, 
those lying in the $x-z$ plane were found to be most important, 
in the sense that they exist and are stable in models 
with the widest variety of cusp slopes and axis ratios.
In particular, the $x-z$ bananas are unstable only in highly 
flattened, nearly prolate models, and they are often more
elongated than the model figure, making them useful 
building-blocks for a self-consistent galaxy.
The boxlets in the $x-y$ and $y-z$ planes are almost always 
vertically unstable in strongly triaxial models (when they exist), 
and it seems unlikely that they could play an important role in 
maintaining triaxiality in real galaxies.

We have investigated only motion in the principal planes.
Other resonant orbits -- although generally of a higher order -- 
exist outside of the principal planes.
Merritt \& Fridman (1996) cataloged the orbits in a single triaxial 
model with Dehnen's density law, $\gamma=1$, $c/a=0.5$ and 
$T=0.5$.
Their Figure 4 confirms that only the $x-z$ banana and the $x-z$ 
fish, among the closed boxlets investigated here, are generators of 
regular orbit families.
The $x-y$ pretzel was also found to be important at low energies, 
as were a number of higher-order resonances outside of the principal planes, 
including the $4:5:7$, $5:6:8$, and $6:7:9$ closed boxlets.

Although instability does not imply stochasticity, unstable 
resonant orbits are often associated with chaos.
We expect that the phase space surrounding many of the unstable 
boxlets is stochastic.
However, the timescale over which the stochasticity manifests 
itself in the orbital motion is likely to be a strong function of 
the model parameters, particularly the cusp slope $\gamma$ and 
the energy.
We will return to this question in a future study.

\bigskip

This work was supported by NSF grant AST 90-16515 and by NASA 
grant NAG 5-2803.

\clearpage

\figcaption[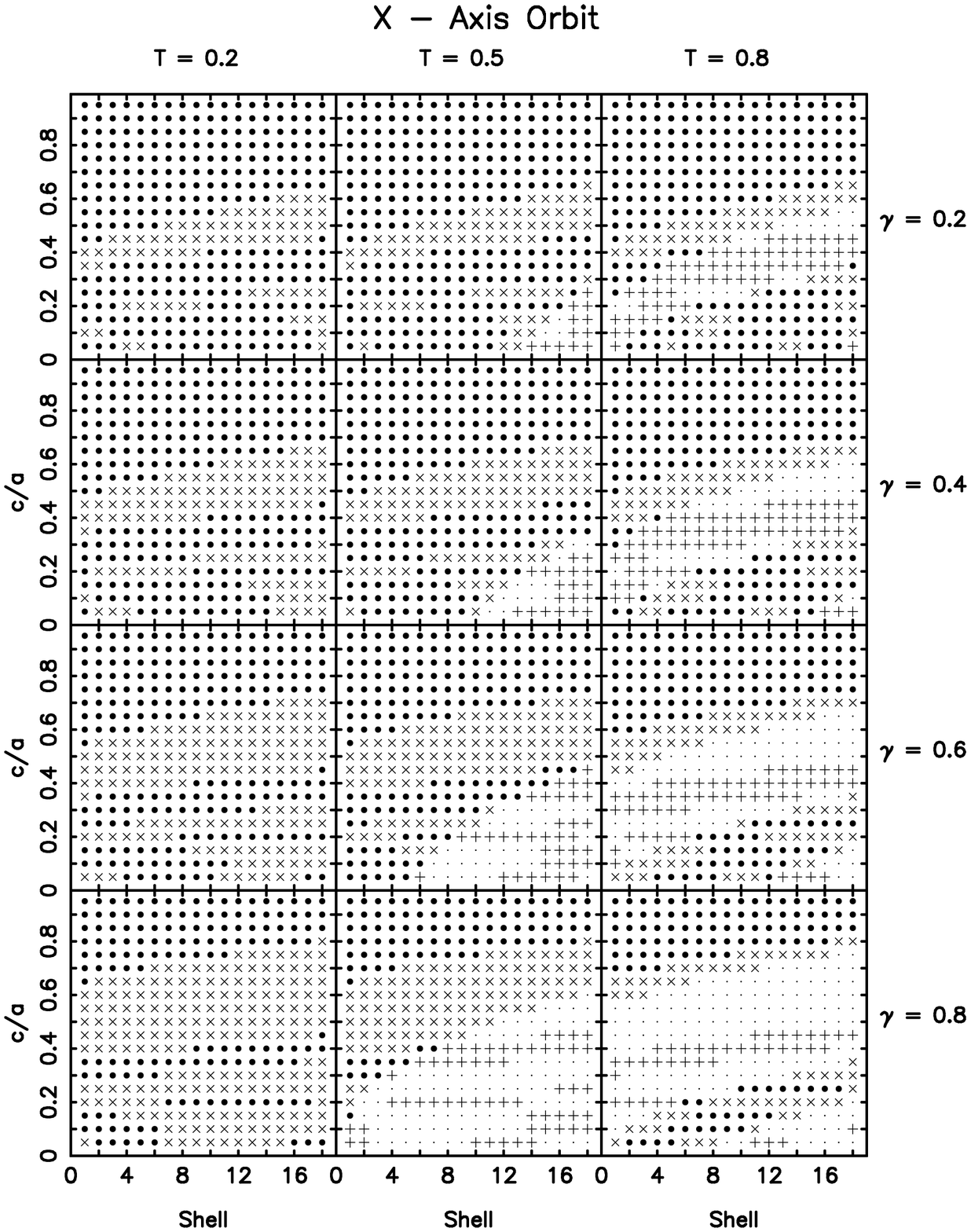]{\label{fig1}} 
Stability diagram for the $x$-axis orbit.
$\bullet$ stable; $+$ unstable in the direction of the $y$ axis;
$\times$ unstable in the direction of the $z$ axis; $\cdot$ 
unstable in both directions. 

\figcaption[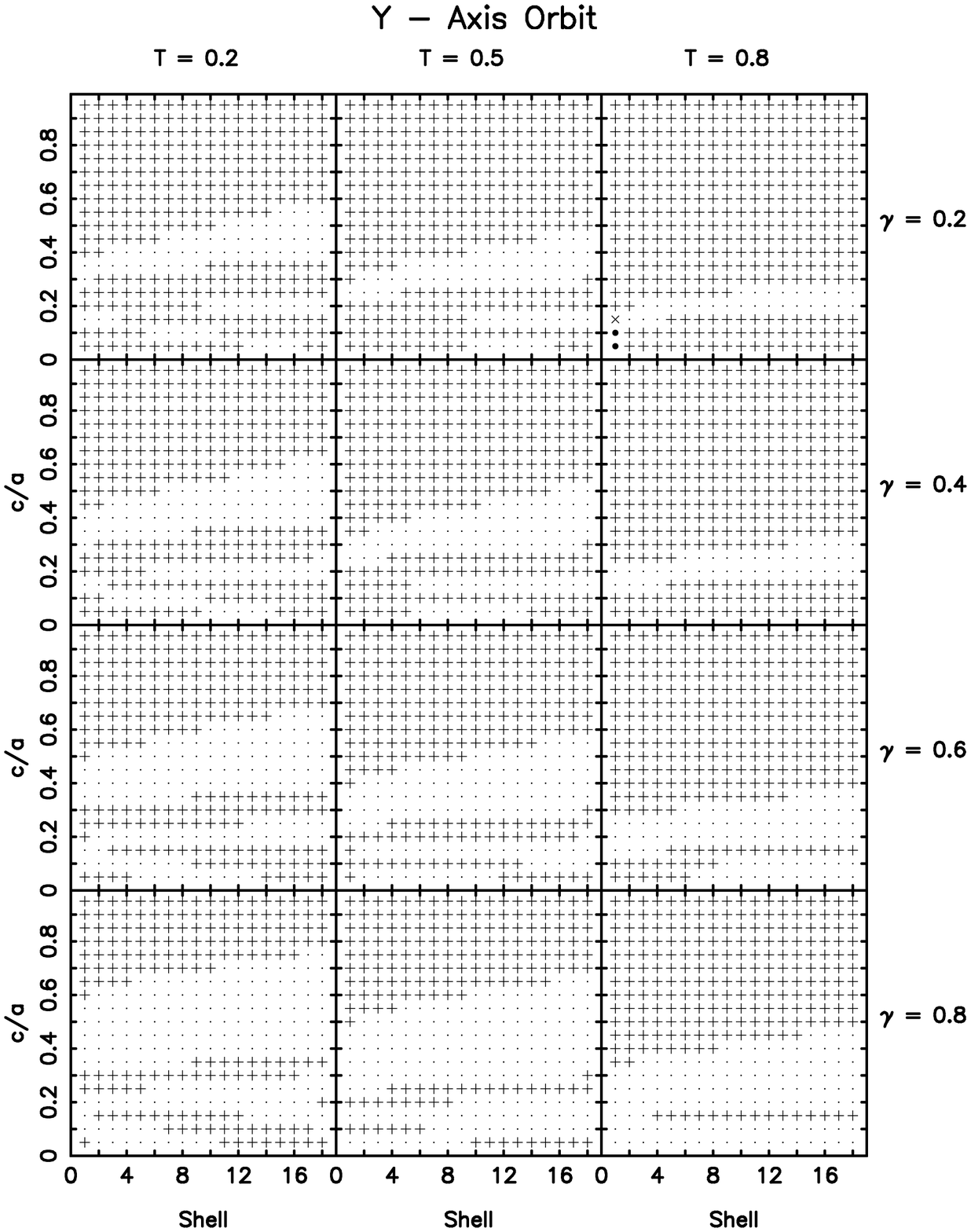]{\label{fig2}} 
Stability diagram for the $y$-axis orbit.
$+$ unstable in the direction of the $x$ axis;
$\times$ unstable in the direction of the $z$ axis; $\cdot$ 
unstable in both directions. 

\figcaption[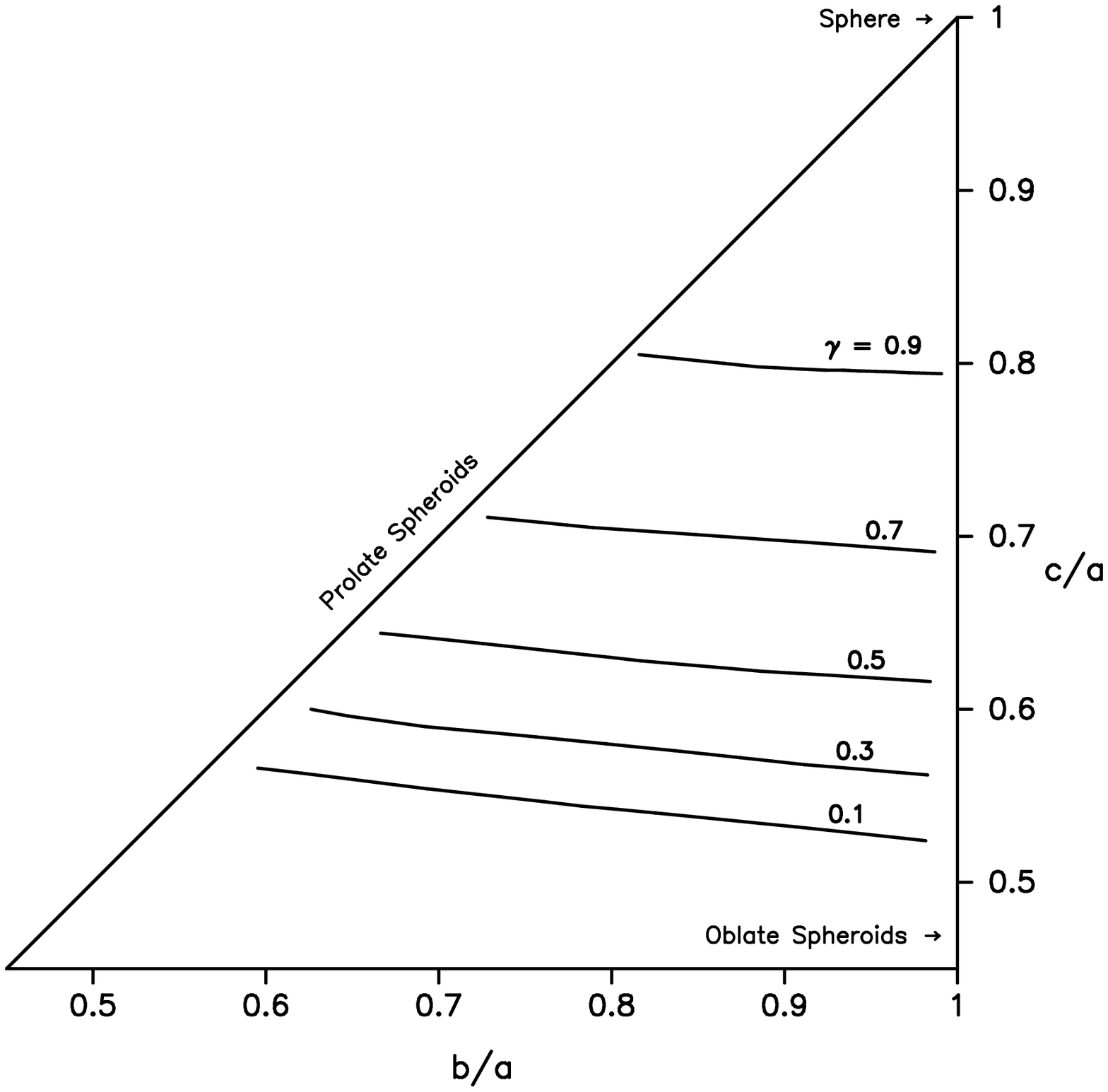]{\label{fig3}} 
Stability region for the $x$-axis orbit.
Solid lines correspond to models in which 
the $x$-axis orbit is just unstable, at the half-mass 
radius (shell 10), to bifurcation of the $x-z$ banana orbit.
Box orbits do not exist outside the half-mass radius in models
below these lines. 

\figcaption[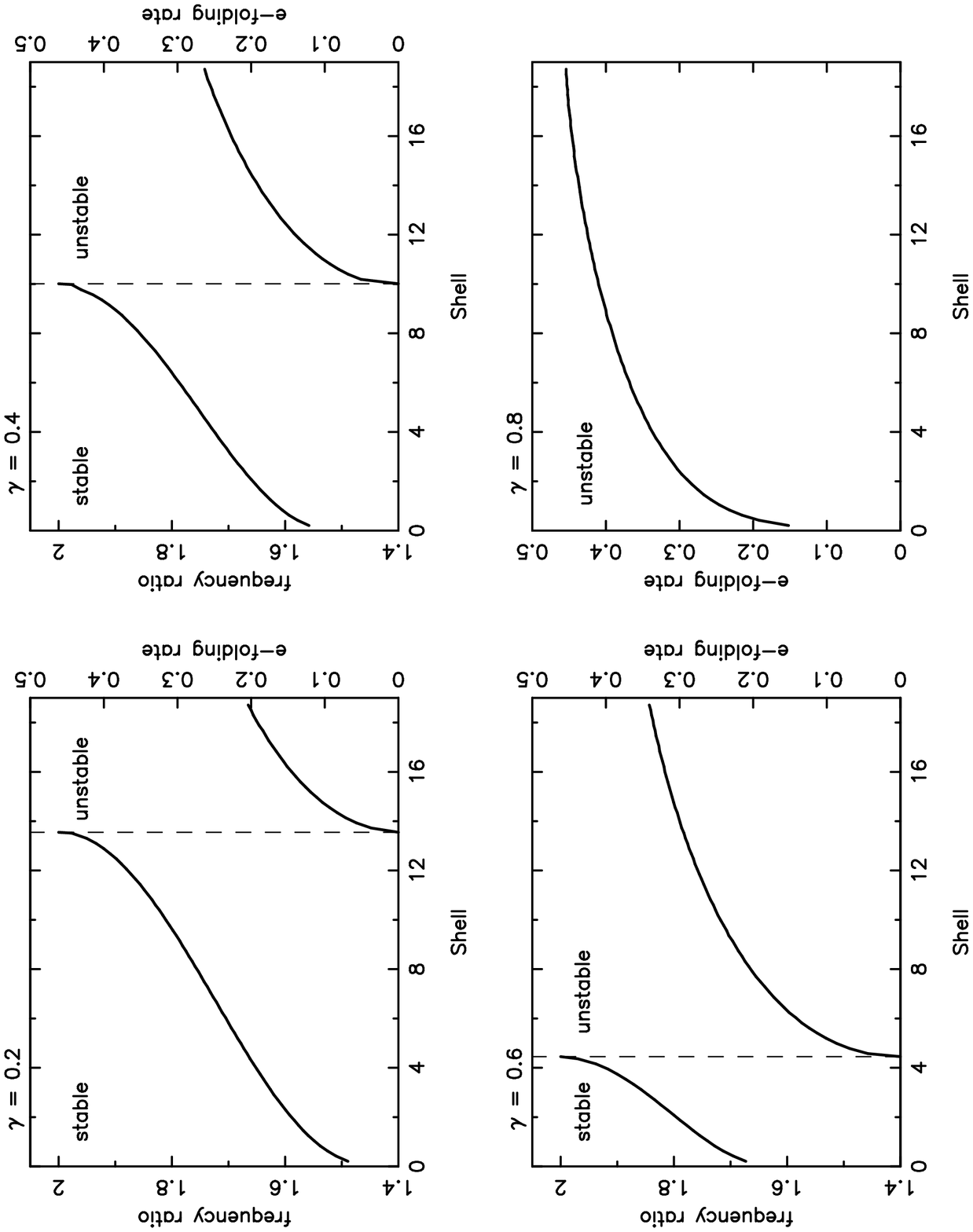]{\label{fig4}} 
Stability characteristics of the $x$-axis orbit when perturbed 
in the $z$-direction, for $c/a=0.6$ and $T=0.5$.
Frequencies and e-folding rates are expressed in units of 
the orbital frequency.
Bifurcation of the $x-z$ banana orbit occurs at successively 
lower energies as the cusp slope $\gamma$ is increased.

\figcaption[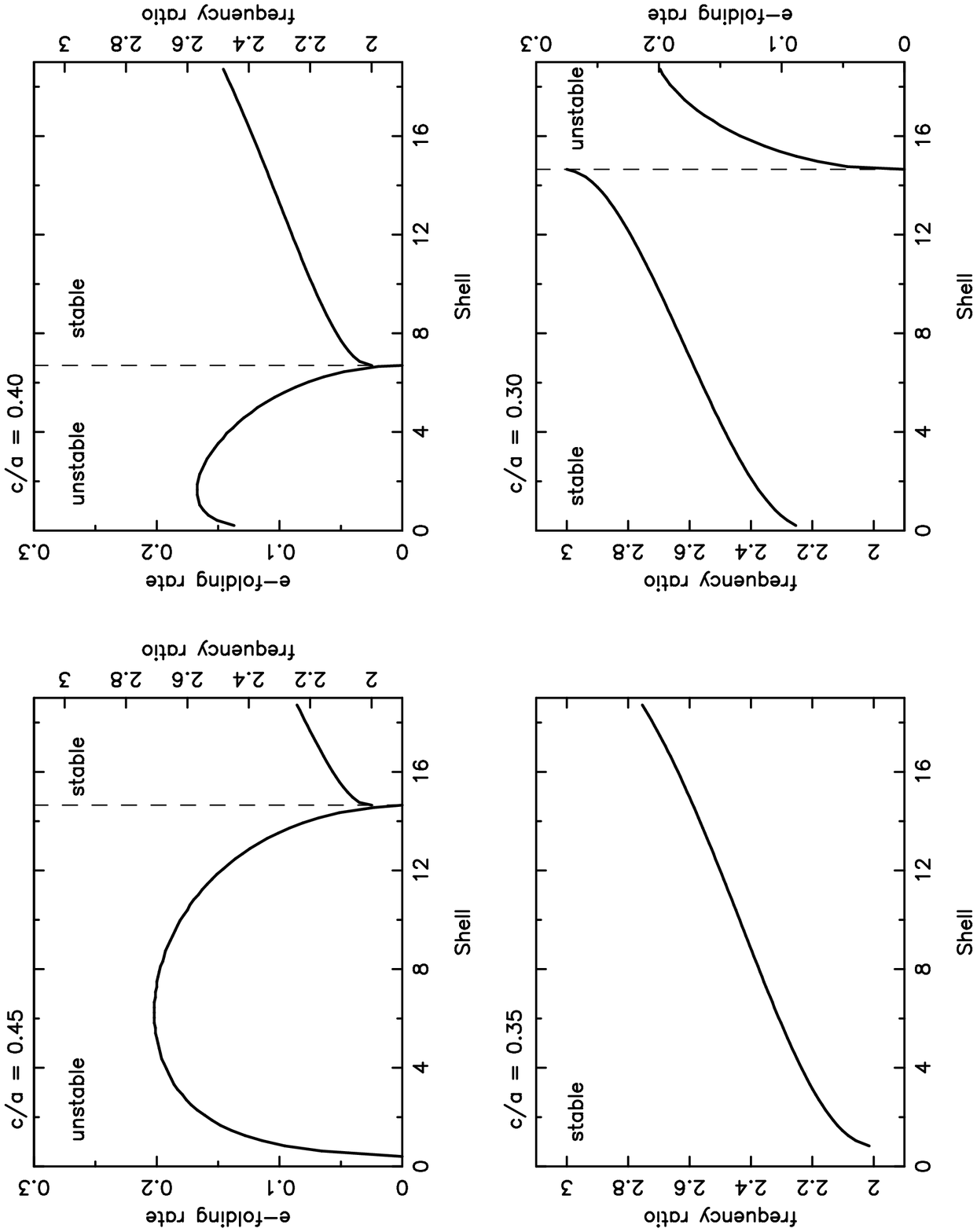]{\label{fig5}} 
Stability characteristics of the $x$-axis orbit when perturbed 
in the $z$-direction, for $\gamma=0.4$ and $T=0.5$.
Frequencies and e-folding rates are expressed in units of 
the orbital frequency.
The $x$-axis orbit returns to stability through bifurcation of the 
$x-z$ anti-banana orbit.

\figcaption[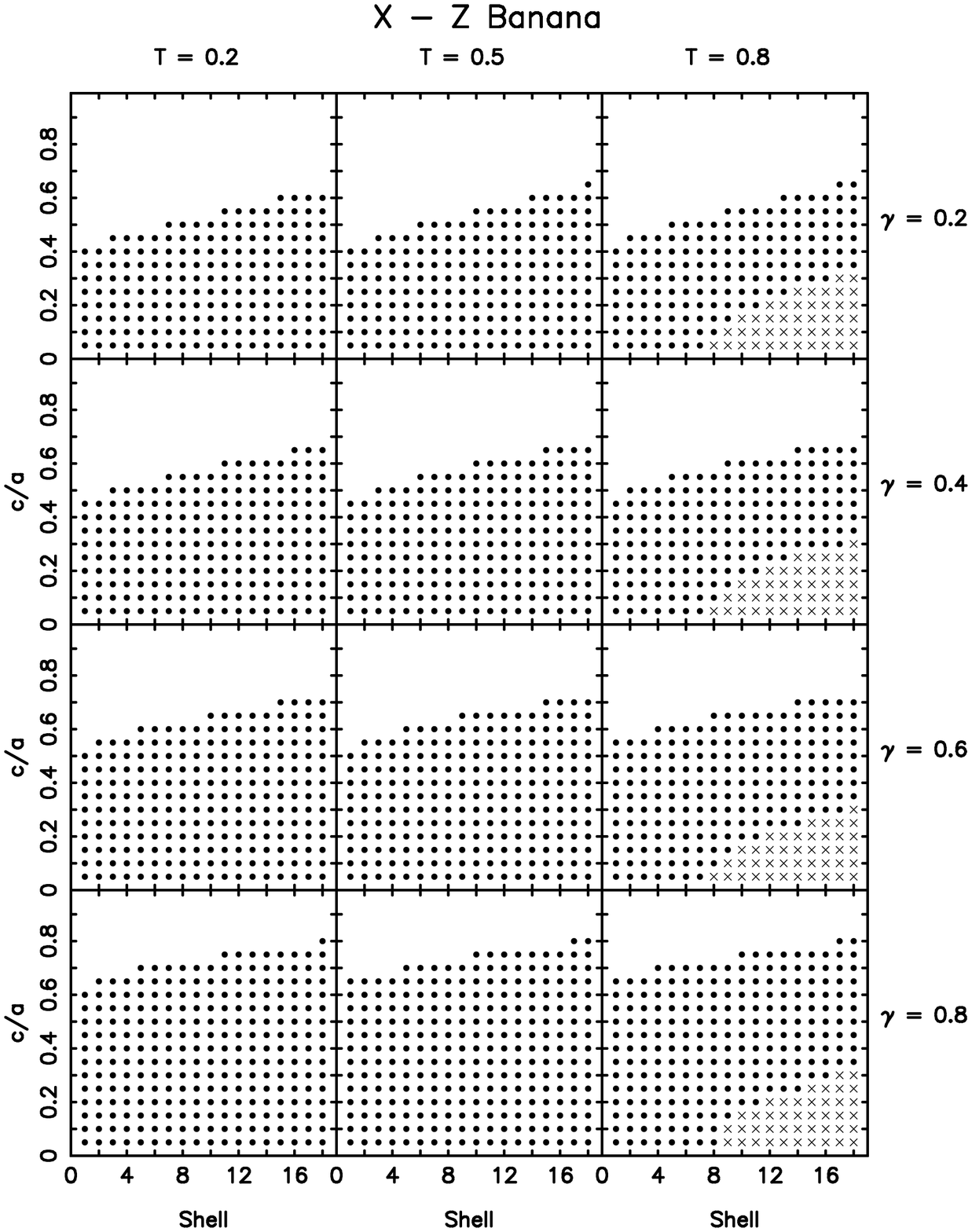]{\label{fig6}} 
Stability diagram for the $x-z$ banana orbit.
$\bullet$ stable; $\times$ vertically unstable.
No symbol indicates that the $x-z$ banana does not exist.

\figcaption[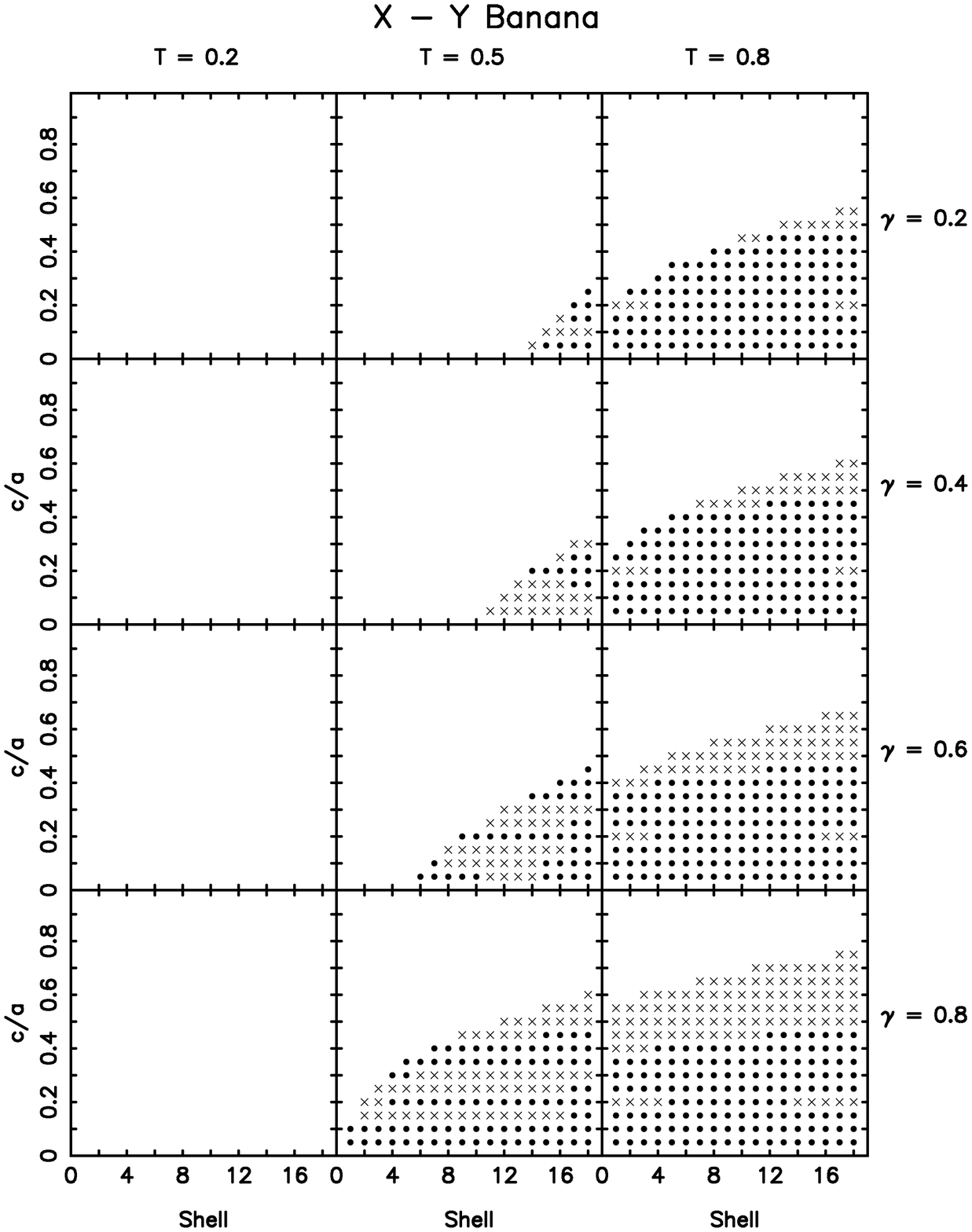]{\label{fig7}} 
Stability diagram for the $x-y$ banana orbit.
$\bullet$ stable; $\times$ vertically unstable.
No symbol indicates that the $x-y$ banana does not exist.

\figcaption[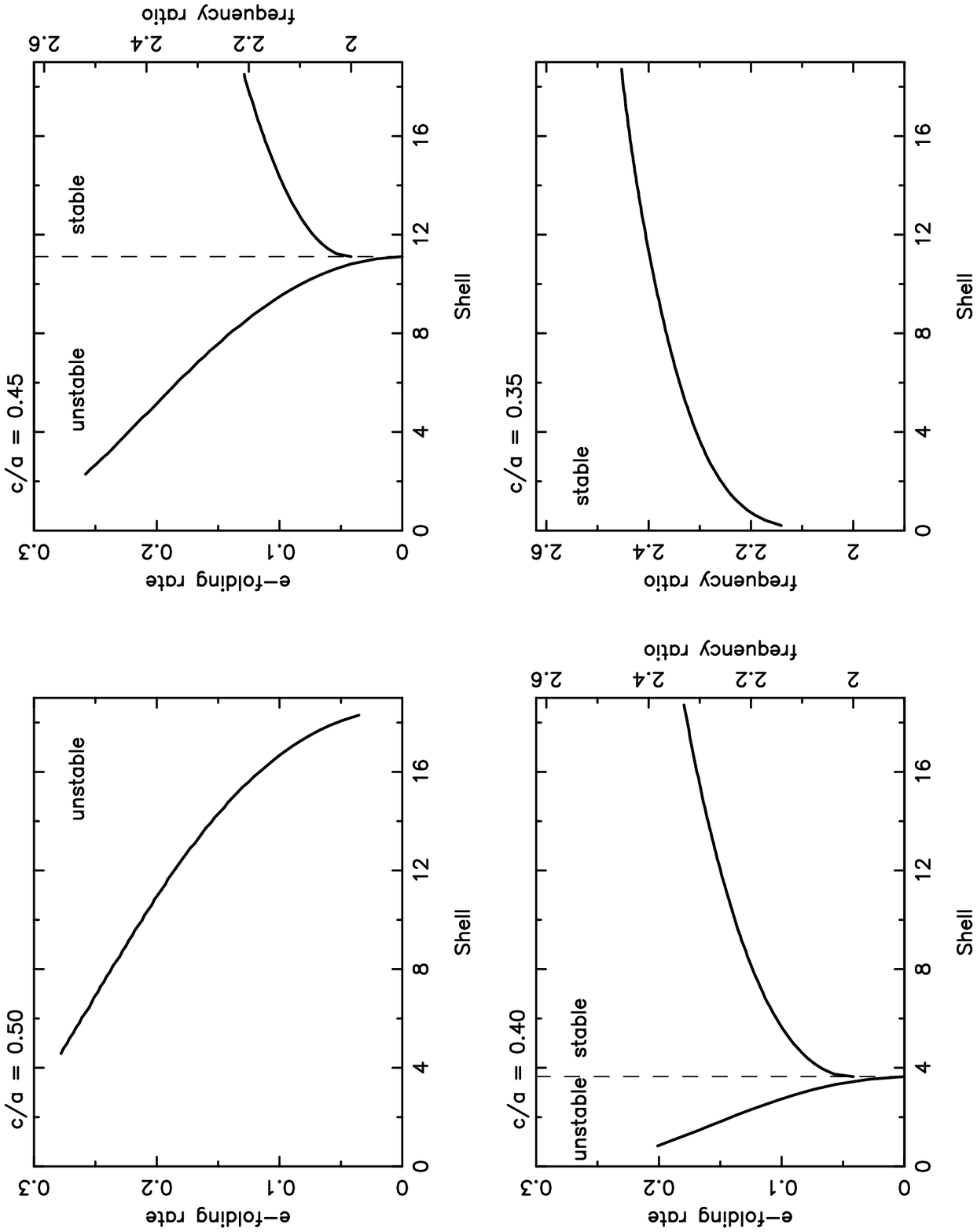]{\label{fig8}} 
Stability characteristics of the $x-y$ banana orbit when perturbed 
in the $z$-direction, for $\gamma=0.6$ and $T=0.8$.
Frequencies and e-folding rates are expressed in units of 
the orbital frequency.

\figcaption[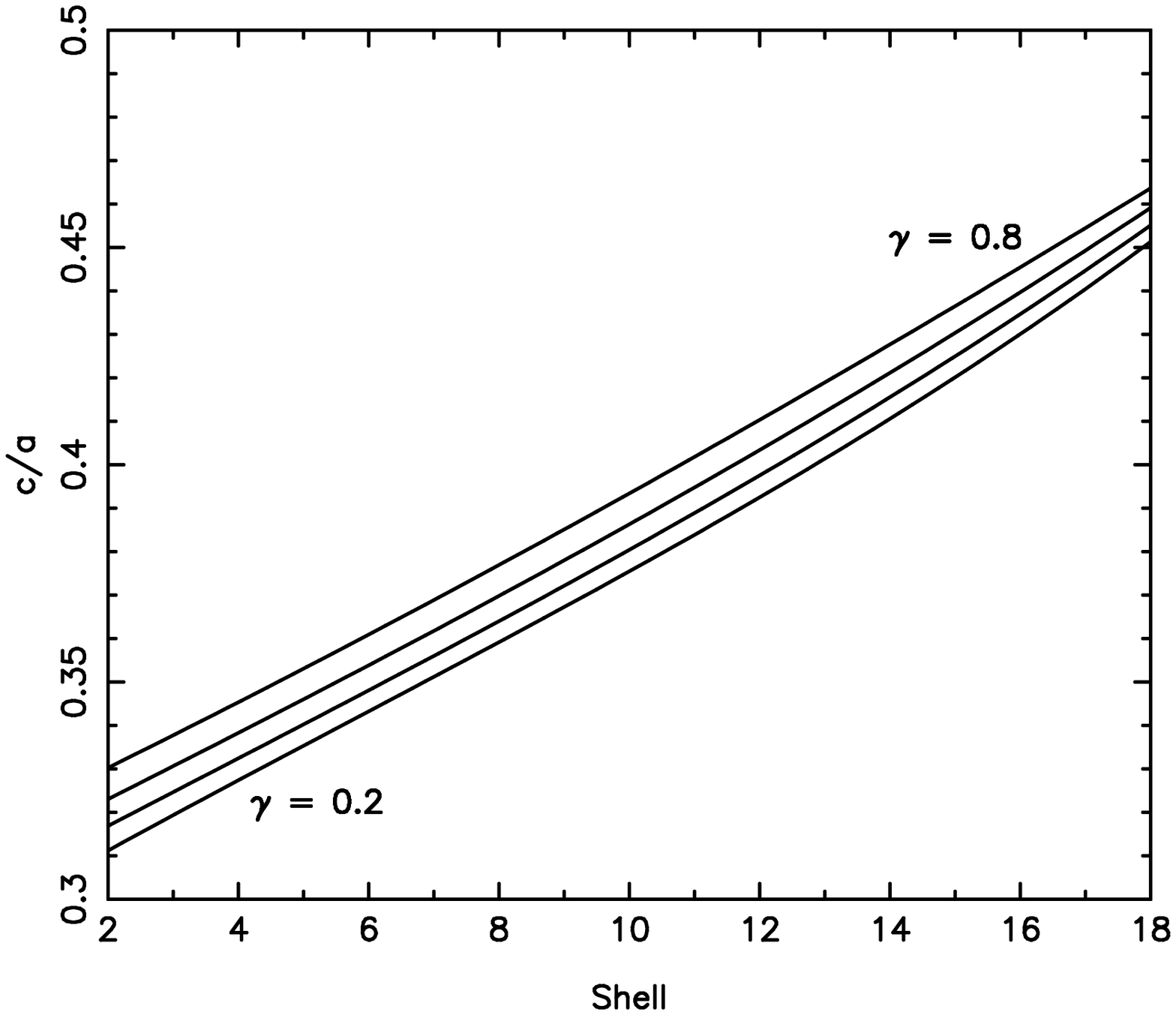]{\label{fig9}} 
Value of $c/a$ at which the elongation of the $x-z$ banana 
matches that of the model, for $T=0.5$, and for four values of 
the cusp slope: $\gamma = 0.2,0.4,0.6,0.8$.

\figcaption[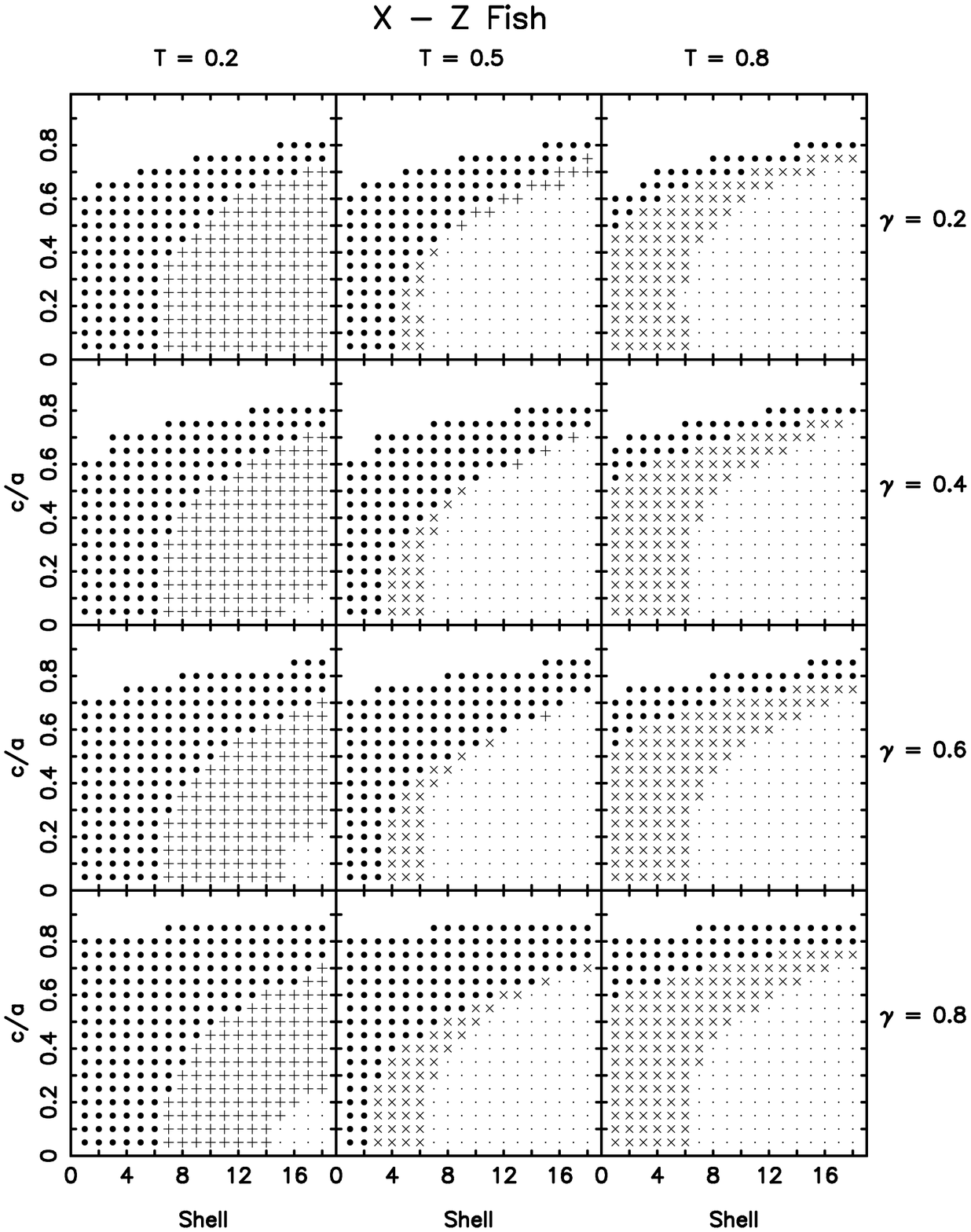]{\label{fig10}} 
Stability diagram for the $x-z$ fish orbit.
$\bullet$ stable; $+$ unstable in the plane; 
$\times$ vertically unstable.
No symbol indicates that the $x-z$ fish does not exist.

\figcaption[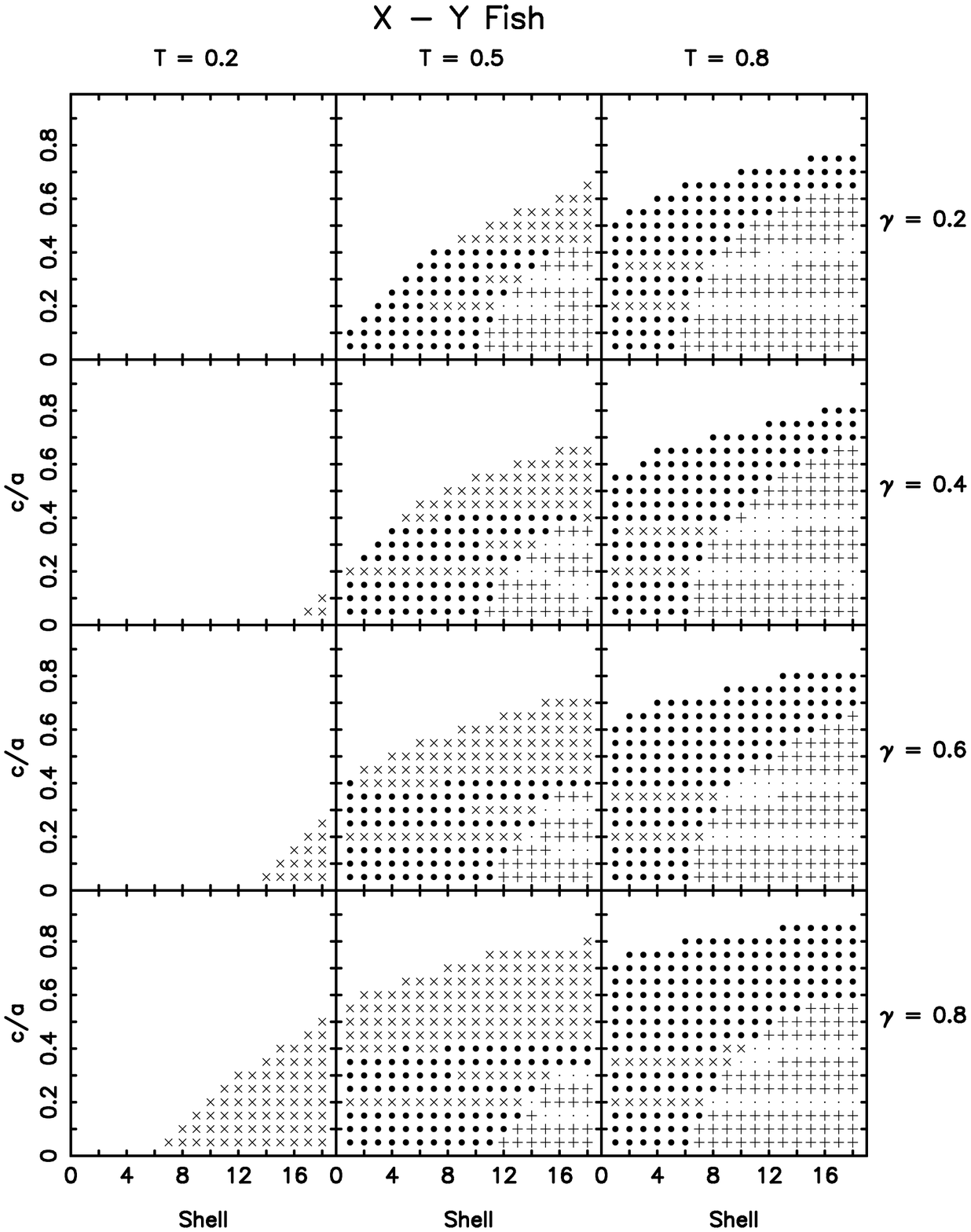]{\label{fig11}} 
Stability diagram for the $x-y$ fish orbit.
$\bullet$ stable; $+$ unstable in the plane; 
$\times$ vertically unstable.
No symbol indicates that the $x-y$ fish does not exist.

\figcaption[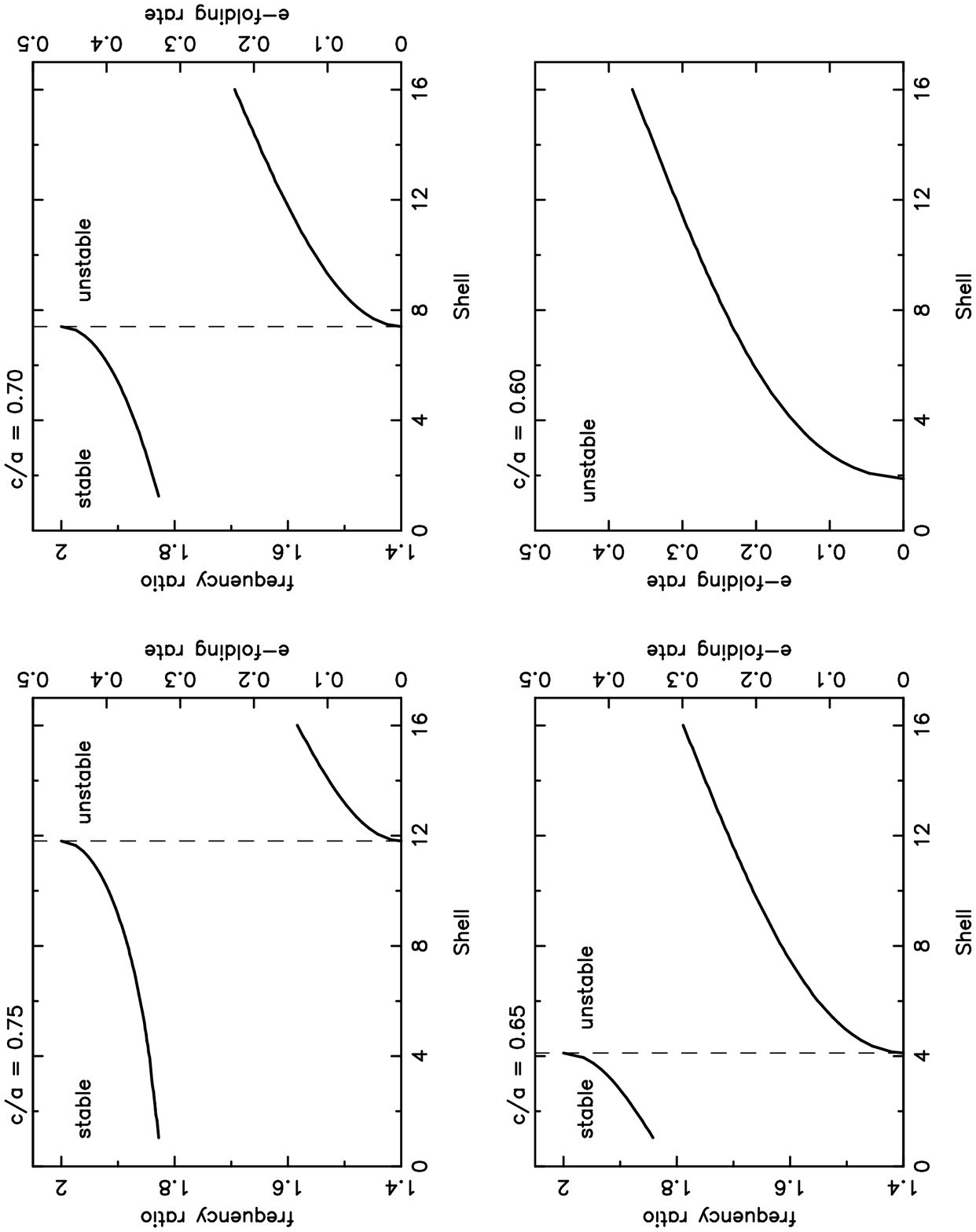]{\label{fig12}} 
Stability characteristics of the $x-z$ fish orbit when perturbed 
in the $y$-direction, for $\gamma=0.8$ and $T=0.8$.
Frequencies and e-folding rates are expressed in units of 
the orbital frequency.

\setcounter{figure}{0}

\begin{figure}
\plotone{figure1.ps}
\caption{ }
\end{figure}

\begin{figure}
\plotone{figure2.ps}
\caption{ }
\end{figure}

\begin{figure}
\plotone{figure3.ps}
\caption{ }
\end{figure}

\begin{figure}
\plotone{figure4.ps}
\caption{ }
\end{figure}

\begin{figure}
\plotone{figure5.ps}
\caption{ }
\end{figure}

\begin{figure}
\plotone{figure6.ps}
\caption{ }
\end{figure}

\begin{figure}
\plotone{figure7.ps}
\caption{ }
\end{figure}

\begin{figure}
\plotone{figure8.ps}
\caption{ }
\end{figure}

\begin{figure}
\plotone{figure9.ps}
\caption{ }
\end{figure}

\begin{figure}
\plotone{figure10.ps}
\caption{ }
\end{figure}

\begin{figure}
\plotone{figure11.ps}
\caption{ }
\end{figure}

\begin{figure}
\plotone{figure12.ps}
\caption{ }
\end{figure}

\end{document}